\documentclass[aps,prb,twocolumn,10pt,showpacs,amsmath,amssymb,floatfix]{revtex4-1}
\usepackage{amsmath}
\usepackage{graphicx}
\usepackage{verbatim}
\usepackage{color}
\usepackage{subfigure}
\usepackage{hyperref}
\usepackage{soul}
\usepackage{amssymb}
\usepackage{cancel}
\usepackage{float}
\begin{document}
%
\title{Chain-like transitions in Wigner crystals: Sequential or non-sequential?}

\author{J. E. \surname{Galv\'an-Moya}} \email[Email: ]{edogalvan@gmail.com} 
\affiliation{Department of Physics, University of Antwerp, Groenenborgerlaan 171, B-2020, Antwerp, Belgium}

\author{V. R. \surname{Misko}} \email[Email: ]{vyacheslav.misko@uantwerpen.be}
\affiliation{Department of Physics, University of Antwerp, Groenenborgerlaan 171, B-2020, Antwerp, Belgium}

\author{F. M. \surname{Peeters}} \email[Email: ]{francois.peeters@uantwerpen.be}
\affiliation{Department of Physics, University of Antwerp, Groenenborgerlaan 171, B-2020, Antwerp, Belgium}

%
\begin{abstract}
The structural transitions of the ground state of a system of repulsively interacting particles
 confined in a quasi-one-dimensional channel, and the effect of the interparticle interaction as
 well as the functional form of the confinement potential on those transitions are investigated.
 Although the non-sequential ordering of transitions (non-SOT), i.e. $1$-$2$-$4$-$3$-$4$-$5$-$6$-$...$
 sequence of chain configurations with increasing density, is widely robust as predicted in a
 number of theoretical studies, the sequential ordering of transitions (SOT),
 i.e.~$1$-$2$-$3$-$4$-$5$-$6$-$...$ chain, is found as the ground state for long-ranged
 interparticle interaction and hard-wall-like confinement potentials.
 We found an energy barrier between every two different phases around its transition point,
 which plays an important role in the preference of the system to follow either a SOT or a non-SOT.
 However, that preferential transition requires also the stability of the phases during the transition.
 Additionally, we analyze the effect of a small structural disorder on the transition
 between the two phases around its transition point.
 Our results show that a small deformation of the triangular structure, change dramatically the
 picture of the transition between two phases, removing in a considerable region the non-SOT in the
 system.
 This feature could explain the fact that the non-SOT is, up to now, not observed in experimental
 systems, and suggests a more advanced experimental set-up to detect the non-SOT.
\end{abstract}
\maketitle

\section{Introduction}

At low temperatures, a classical system of charged particles arranges itself in a close packet
 structure, also known as Wigner crystal\cite{045_wigner, 190_shikin, 182_grimes}.
 This organization of particles allows the lowest energy configuration and, due to the absence
 of kinetic energy, the arrangement of particles results in a stable crystalline structure.
 In a two-dimensional system, Wigner crystals have a hexagonal lattice
 structure.\cite{004_schweigert,009_partoens,040_meyer}.

For a quasi-one-dimensional (Q1D) system of classical charged particles confined in a parabolic
 channel, Piacente \emph{et al.}\cite{165_piacente,*003_piacente} predicted a non-sequential
 ordering of transitions (non-SOT) between ground state (GS) configurations:
 $1$-$2$-$4$-$3$-$4$-$5$-$6$-$...$ chain-like structures with increasing particle density.
 They found that the range of the interaction between particles does not affect the
 ordering of the transition, when the particles are confined by a parabolic
 potential\cite{173_piacente,006_piacente,041_galvan}.
 The structural transition from two- to four-chain configuration ($2 \rightarrow 4$) occurs, in
 the case of a non-SOT, through a zigzag transition of each of the two chains and a simultaneous
 small shift along the chain, which makes it a discontinuous transition\cite{003_piacente}.

Although this non-SOT has been found as the GS in a number of theoretical
 works\cite{165_piacente,003_piacente,147_koppl,171_klironomos}, it was not, to the best of our
 knowledge, observed in experiments so far.
 Instead, in the experiments they observed a direct transition
 from two- to three-chain configuration ($2 \rightarrow 3$), allowing the system the follow a
 usual sequential order of transitions (SOT), with increasing linear density, as reported in the
 case of  electrons on liquid He at low temperatures\cite{169_ikegami, 213_rees, 214_rees} and
 even in dusty plasma clusters~\cite{037_sheridan}.
 The SOT in a system of electrons on liquid He has been indirectly measured by interpreting the
 step-like increment of the conductance of a channel, while an electrical force guides the
 motion for the particles in the structure\cite{169_ikegami, 213_rees}.
 Theoretical works have modeled this system, evidencing the SOT during that dynamical
 process\cite{215_araki,188_vasylenko,216_vasylenko},
 although modifying the shape (i.e., increasing the length) of the confining constriction was
 shown to facilitate the observations of the non-SOT\cite{216_vasylenko}.
 Similarly, a SOT has been predicted theoretically for a binary mixture of repulsive particles
 trapped in a channel~\cite{012_ferreira}, and for an Abrikosov-vortex arrangement in a
 superconducting slab for low
 temperatures~\cite{193_guimpel,194_brongersma,195_carneiro,192_sardella,172_barba}, as well as
 for Pearl vortices~\cite{134_bronson}.
 
Ikegami~\emph{et al.}\cite{169_ikegami} observed a periodic change of melted and ordered states as
 a function of linear density, i.e. re-entrant melting. However, in that work melting temperature
 was just a measure of disorder of the system.  In terms of structural transitions, they always
 observed SOT, in agreement with Ref.~[\onlinecite{003_piacente}] if temperature is above a
 certain value.  However, following the results of Piacente~\emph{et al.}\cite{003_piacente}
 one might expect that lowering temperature would result in the non-SOT, which was not seen
 experimentally within the attainable temperature range\cite{169_ikegami, 213_rees}.
 Thus the important question rises: how to optimize experimental set-ups in order to realize the
 observation of the non-SOT?
 
Previous theoretical works have shown the robustness of the non-SOT for a system of particles
 trapped in a perfect parabolic channel irrespective of the range of the interaction between the
 particles\cite{003_piacente}, and also for a system of particles with a fixed interparticle
 interaction, confined in Q1D channels with potentials of various functional form.\cite{211_galvan} 
 However, it has been recently proven that it is possible to find a SOT region by tuning
 the parameters of the confinement potential, e.g. by invoking a profile similar to the
 Bean-Livingston barrier for vortices in a superconductor\cite{211_galvan}.
 At first glance, we note that two extreme cases of this problem have been analyzed: a fixed
 confinement varying the range of the interaction, and a fixed interaction range varying the
 profile of the confinement channel.
 From these results a trend of the ordering of the transitions was revealed, but it still did not
 resolve two important issues: why the system prefers following a non-SOT rather than SOT?
 And, most importantly, why experiments did not find a non-SOT if it is a {\it preferable} scenario?
 In this work, we present an in-depth analysis which answers these questions.

The present paper is organized as follows. We first give, in Sec.~\ref{sec_model}, an overview
 of the model and the perspective of the current work done in this area. Sec.~\ref{sec_range} is
 devoted to the analytical study of the effect of the interaction range on the ordering of
 the structural transitions, for different confinement potentials. In Sec.~\ref{sec_barrier}, we
 induce transitions from two- to three-chain ($2 \rightarrow 3$) and from two- to four-chain
 ($2 \rightarrow 4$), and calculate the energy barrier for each transition, analyzing the
 stability of the configurations formed in the vicinity of the transition point.
 The effect of small imperfections on the energy barrier and on the ordering of the transitions
 are analyzed in Sec.~\ref{sec_imperfections}.
 Finally, our conclusions are given in Sec.~\ref{conclusions}

\section{Model System} \label{sec_model}

First, we consider a two-dimensional infinite system of identical interacting particles with
 electric charge $q$ and mass $m$, which move in the $xy$-plane. The particles are confined by
 a one-dimensional potential limiting their motion in the $y$-direction, forming a
 quasi-one-dimensional channel along the $x$-axis [$V_{conf}(y)$].
 The total energy of the system is given by:
\begin{equation}
 H = \sum_{i=1}^{\infty} \sum_{j>i}^{\infty}
	    V_{int}(|\mathbf r_{i} - \mathbf r_{j}|) + \sum_{i=1}^{\infty} V_{conf}(y_{i}),
\end{equation}
 where $\mathbf r_{i}$ is the relative position of the $i$-th particle in the system, while
 $V_{int}(r)$ represents the pairwise inter-particle interaction, which is taken as a screened
 power-law potential, which will allow for the simulations of both short- and long-range
 interactions, as follows:
 \begin{eqnarray}
  V_{int}(r) & = & \frac{q}{\epsilon R} \frac{R^n e^{-\kappa r/\lambda}}{r^n},
\end{eqnarray}
 where the parameters $\lambda$ and $n$ allows us to tune the range of the interaction between
 particles in the system, $\epsilon$ is the dielectric constant of the medium the particles are
 moving in, and $R$ is an arbitrary length parameter which we introduced to guarantee the right
 units.
 
Due to the importance of the profile of the confinement potential on the ordering of the
 transitions, as shown in Ref.~[\onlinecite{211_galvan}], we use the following two different
 functional form of the confinement potential, which allows us to vary the profile of the channel
 continuously from a parabolic-like to a hard-wall potential:
 \begin{eqnarray}
  V_{A}(\beta,y)  & = & \frac{m\upsilon_A^2 y_{0}^{2}}{2}
			  \frac{\cosh(\beta y) - 1}{\cosh(\beta y_0)-1}, \label{conf_cosh_orig} \\
  V_{B}(\gamma,y) & = & \frac{m\upsilon_B^2 y_{0}^{2}}{2}
			  \left[ \text{e}^{-\gamma^2(y-y_{0})^2}
			       + \text{e}^{-\gamma^2(y+y_{0})^2} \label{conf_gauss_orig} \right],
\end{eqnarray}
where $\beta$ and $\gamma$ control the sharpness, $\upsilon$ the strength, and $y_0$ the effective
 width of the confining channel. In the following, we refer to $V_{A}(\beta,y)$ [$V_{B}(\gamma,y)$]
 as exponential [Gaussian] confinement.
 
In dimensionless form, the interaction and the confinement potentials in our model become:
\begin{eqnarray}
  V_{int}(r) & = & \frac{e^{-\kappa r}}{r^n}, \label{Vint}\\
  V_{A}(\beta,y)  & = & \upsilon^2 y_{0}^{2}
			    \frac{\cosh(\beta y) - 1}{\cosh(\beta y_0)-1}, \label{conf_cosh} \\
  V_{B}(\gamma,y) & = & \sigma^2 y_{0}^{2}
			      \left[ \text{e}^{-\gamma^2(y-y_{0})^2}
				   + \text{e}^{-\gamma^2(y+y_{0})^2} \label{conf_gauss} \right],
\end{eqnarray}
where the energy is expressed in units of
 $E_0 = (m\omega_0^2/2)^{n/(n+2)} (q^2/\epsilon)^{2/(n+2)} R^{2(n-1)/(n+2)}$ and all distances
 are expressed in units of $r_0 = (2q^2/m\omega_0^2\epsilon)^{1/(n+2)} R^{(n-1)/(n+2)}$. 
 The dimensionless frequencies are given by $\upsilon=\upsilon_A/\omega_0$ and
 $\sigma=\upsilon_B/\omega_0$, while $\omega_0$ measures the strength of the confinement
 potential, and the screening of the pairwise interaction is $\kappa=r_0/\lambda$.  The
 dimensionless linear density $\eta$ is defined as the number of particles per unit of length
 along the unconfined direction.

Concerning the sequence of the GS configuration for increasing system density, previous works have
 shown that:
 1)~In case of parabolic confinement, the GS follows the non-SOT irrespective of the range of the
 interparticle interaction\cite{003_piacente}.
 2)~In case of the interaction potential given by $n=1$ and $\kappa=1$ in Eq.~(\ref{Vint}), and
 varying the profile of the channel, the GS follows the non-SOT in the limiting cases of hard-wall
 and parabolic-like profile\cite{211_galvan}. The latter study also showed that non-SOT is present
 in all systems for intermediate values of the shape parameter, evidencing that the non-SOT is
 extremely robust for a broad range of possible profiles and shape parameters.  However, in case
 of Gaussian confinement, SOT for the GS was found within a small window of the shape
 parameters\cite{211_galvan}, indicating that the shape of the channel profile is of crucial
 importance when one is looking for a SOT in the system.
  
These results bring in evidence the strong relation between the range of the interparticle
 interaction and the confinement profile.  Due to the fact
 that in experiments it is very common to observe the SOT but not the non-SOT, a complete and
 detailed study about the effects of the range of the interparticle interaction and the profile of
 the confinement channel on the ordering of the transitions, are needed, as well as a
 trustable recipe for the experimentalist about the regions where the non-SOT could be observed.
 The present paper will address those problems.

\section{Interaction range vs. confinement profile} \label{sec_range}
\begin{figure*} [t!]
\begin{center}
\subfigure{\includegraphics[trim={0.7cm 0 0 0}, scale=0.48]{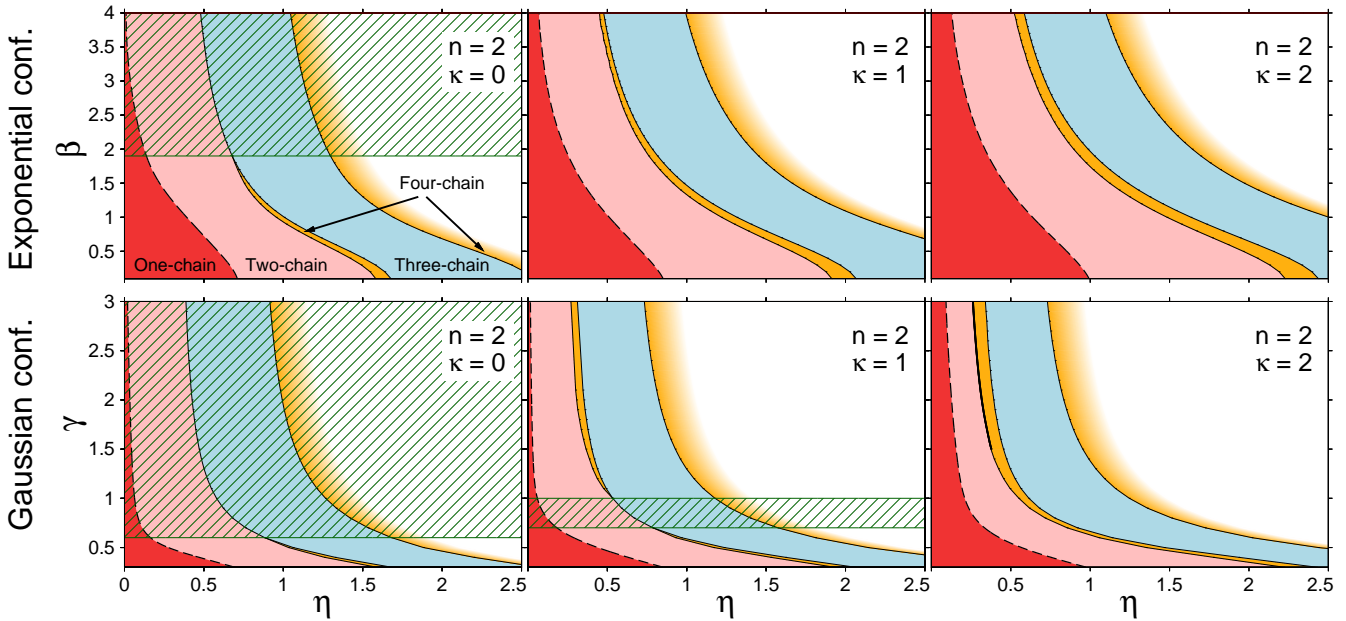}}
\caption{\label{fig:phdiag_nk} (Color online) Phase diagrams of the ground state for exponential
 (upper pannel) and Gaussian confinement (lower panel), for different interaction potentials as
 indicated in the figure.  Each phase diagram shows the GS as function of shape parameter ($\beta$
 or $\gamma$) and linear density of the system ($\eta$). Solid and dashed lines represent first
 and second order transitions, respectively.  Regions highlighted by a line pattern indicate the
 existence of SOT in the GS of the system.}
\end{center}
\end{figure*}

Analytical calculation of the energy of a system of different chain-like structures, is performed by
 following the model proposed in Ref.~[\onlinecite{211_galvan}].  As a result, we found
 "shape parameter vs. density" phase diagrams for different confinement potentials and
 different ranges of the interaction between particles, as shown in Fig.~\ref{fig:phdiag_nk}. 
 The phase diagrams are the zero temperature GS configuration as function of the shape parameters and the 
 linear particle density $\eta$. All the transitions are of first order, except the zigzag transition
 between one- and two-chain which is of second order.  Interestingly, these results show the
 appearance of a region where the GS is guided by a SOT, for three of the model systems, as
 highlighted by the  patterned region in each phase diagram.
 These results evidence the complementing behavior between the confinement and the interaction
 potential.
 Previously\cite{211_galvan}, SOT had been reported theoretically only in case of Gaussian
 confinement for the interaction parameters $n=1$ and $\kappa=1$. In that case, the SOT was
 found within a small window of $\gamma$ (including $\gamma=1$ which models the Bean-Livingston
 barrier for superconducting vortices).

\begin{figure} [b!]
\begin{center}
\subfigure{\includegraphics[trim={0.0cm 0 0 0}, scale=0.68]{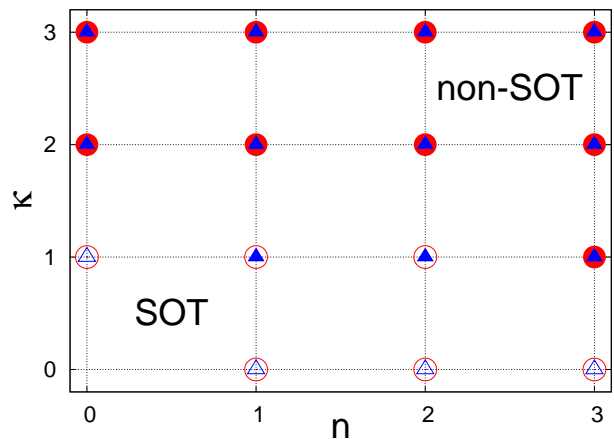}}
\caption{\label{fig:phdiag_sot} (Color online) Extended phase diagram of the ground state
 transition for exponential (blue triangles) and Gaussian confinement (red circles), as a function
 of the interaction parameters $n$ and $\kappa$.
 The solid symbols indicate that GS always follows a non-SOT by increasing density, while open
 symbols indicate the cases where a SOT region has been found.}
\end{center}
\end{figure} 
An extended phase diagram for the ordering of the transitions is plotted in Fig.~\ref{fig:phdiag_sot},
 as a function of the parameters of the interparticle interaction $\kappa$ and $n$, which control the
 range of the interaction (see Eq.~(\ref{Vint})).  The solid symbols
 (triangles for exponential and circles for Gaussian confinement) indicate the case when the GS
 configuration always follows a non-SOT irrespective of the shape parameters, while the open
 symbols indicate the case when the SOT has been found.
 From Fig.~\ref{fig:phdiag_sot} one can see that, the non-SOT is robust for a vast range of shape
 parameters of both confinement potentials studied.
 This behavior allows one to interpret the non-SOT as the "natural" mechanism, which determines
 the behavior of a system of classical particles with increasing particle density,
 irrespective of the interaction between them or the shape of the confinement potential of the
 channel.
 This result is a generalization of previous theoretical
 works\cite{003_piacente,041_galvan,211_galvan} to a large set of different interparticle
 interactions and confinement potentials.

Note that the SOT has been found in some regions of the shape parameters, when the interaction
 potential is long-range. However, the range of the interaction to observe a SOT must be even
 longer in case of exponential confinement as compared to Gaussian confinement.
 This finding is very interesting because the experimental observation of SOT in systems of
 electrons floating on the surface of liquid He, assume a Coulomb-like interaction between
 particles\cite{169_ikegami}, which is clearly a long-range interaction.

\section{Energy barrier at the transition point} \label{sec_barrier}

In order to understand why the system prefers to follow one ordering (SOT versus non-SOT) instead
 of the other one, we compare the energy barrier the system has to overcome during the transitions
 $2 \rightarrow 3$ and $2 \rightarrow 4$.
 
\begin{figure} [t!]
\begin{center}
\subfigure{\includegraphics[trim={0.0cm 0 0 0}, scale=0.68]{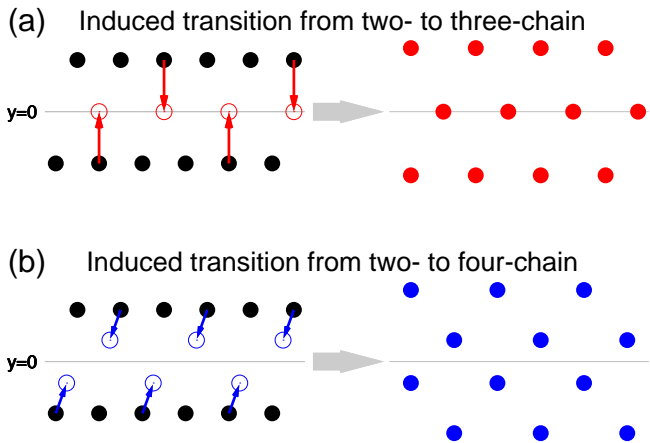}}
\caption{\label{fig:path_transitions} (Color online) Paths of the motion for the dragged particles,
 during the induced transition: a) $2 \rightarrow 3$, and b) $2 \rightarrow 4$.}
\end{center}
\end{figure}

For this purpose, we will focus on the case of the interaction parameters $n=2$ and $\kappa=1$,
 for the Gaussian confinement potential given by Eq.~(\ref{conf_gauss}), because in this case
 the SOT is present in a small window of values of the parameter $\gamma$.
 In order to calculate the energy barrier between the configurations, we set the two-chain structure,
 which was found as metastable or GS configuration close to the transition point ($\eta_t$), as
 initial configuration.
 We induce the transitions $2 \rightarrow 3$ and $2 \rightarrow 4$, at a given density ($\eta_o<\eta_t$), by
 forcing the displacement of some selected particles, allowing the rest of the particles to adjust
 themselves in order to reach the most energetically favorable configuration.
 The selected particles are dragged along straight trajectories as shown in
 Fig.~\ref{fig:path_transitions}(a) and Fig.~\ref{fig:path_transitions}(b) for the transitions
 $2 \rightarrow 3$ and $2 \rightarrow 4$, correspondingly.  The straight paths are expected to be
 close to the real trajectories of the selected particles taking into account the
 symmetry of the initial and the final configurations during the transitions $2 \rightarrow 3$ and
 $2 \rightarrow 4$.
 %

\begin{figure} [b!]
\begin{center}
\subfigure{\includegraphics[trim={0.0cm 0 0 0}, scale=0.68]{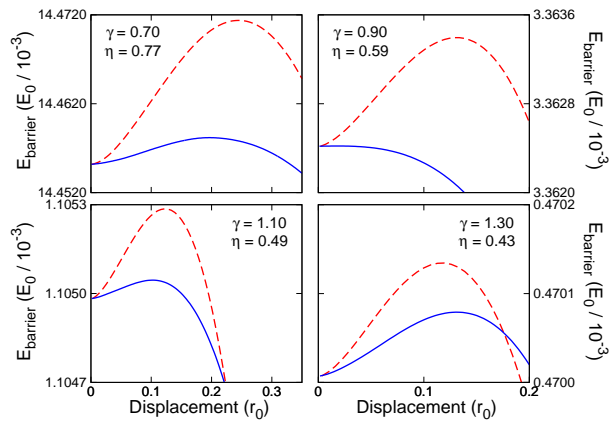}}
\caption{\label{fig:barrier} (Color online) The energy barrier for the induced transitions as a
 function of the displacement of the dragged particles, for different values of the confinement
 parameter and electron density.
 Red dashed line represents the transition $2 \rightarrow 3$, while blue solid line represents the
 transition $2 \rightarrow 4$.}
\end{center}
\end{figure}

Following the displacements shown in Fig.~\ref{fig:path_transitions}, we found that an energy
 barrier is formed very close to the transition point ($\eta_o \lesssim \eta_t$), and at $\eta_t$,
 the barrier for transition $2 \rightarrow 3$ is always {\it higher} than the barrier for
 transition $2 \rightarrow 4$, as shown in Fig.~\ref{fig:barrier}.  This behavior shows that, in
 terms of energy cost, the system always prefers the transition $2 \rightarrow 4$ rather than
 the transition $2 \rightarrow 3$.
 This result provides an understanding, why the system prefers to follow the non-SOT which appears
 to be a straightforward
 mechanism for the transition from a two-chain to a four-chain configuration, due to the minimum
 energy cost for the system.

\begin{figure} [t!]
\begin{center}
\subfigure{\includegraphics[trim={0.0cm 0 0 0}, scale=0.68]{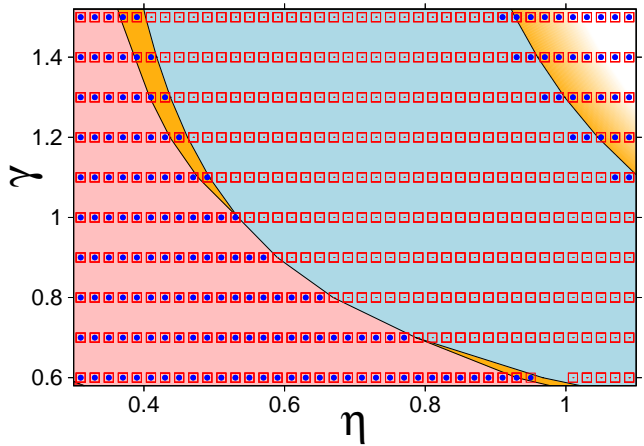}}
\caption{\label{fig:stability} (Color online) Stability region of the transitions $2 \rightarrow 4$
 (solid circles) and $2 \rightarrow 3$ (open squares).
 The regions defined by the different colors,
 represents the analytically calculated phase diagram for the GS, as a function of the shape parameter
 ($\gamma$) and the linear density ($\eta$), as shown in Fig.~\ref{fig:phdiag_nk}.
 We took the parameters $n=2$ and $\kappa=1$.}
\end{center}
\end{figure}

Now, we extend the calculation of the induced transition not only around $\eta_t$, but to the
 whole region of parameters.
 For every displacement during these induced transitions, we calculate the dynamical matrix of the
 system, and by using the Newton optimization\cite{004_schweigert}, we calculate the most
 energetically favorable configurations, together with their vibrational eigenfrequencies.  From
 the symmetry of our system, we know that it has one translational symmetry, which is
 numerically evidenced by the existence of only one vanishing eigenfrequency.  This condition is
 important in our case, due to the fact that it allows us to define the stability of the different
 configurations found, after the induced transition is performed.
 In Fig.~\ref{fig:stability} we show, with symbols, the regions where the induced transitions
 $2 \rightarrow 3$ and $2 \rightarrow 4$ are stable; at the same time we plot the analytical
 results for the phase diagram of the GS (see Fig.~\ref{fig:phdiag_nk}).
 
From Fig.~\ref{fig:stability} one can see that, while the induced transition $2 \rightarrow 3$ is
 stable irrespective of the value of $\gamma$, the stability of the transition $2 \rightarrow 4$ is
 restricted to two different regions of $\gamma$.  We can observe that our numerical results show
 a perfect match with the transition point analytically calculated in Section~\ref{sec_range}.
 It is worth noting that the SOT formed in the window ($0.7 \lesssim \gamma \lesssim 1$), as found
 analytically, arises from the fact that the transition $2 \rightarrow 4$ is not stable at the
 intermediate points ($\gamma=0.8$ and $\gamma=0.9$).
 Summarizing, the system "naturally" evolves following the non-SOT with increasing density, since,
 in all the cases the transition $2 \rightarrow 4$ is more energetically favorable than the
 transition $2 \rightarrow 3$.
 However, the window of non-stability of the configurations during the transition
 $2 \rightarrow 4$, opens the door for the transition $2 \rightarrow 3$ as the GS, and therefore
 the SOT takes place.

\section{The role of fluctuations} 
\label{sec_imperfections}

In the above analysis, we assumed that there is no disorder. 
 As long as a small amount of disorder does not affect the above findings, these can be considered
 as robust and reliable.  
 The goal of this section is to analyze the effect of disorder. 

Typically, disorder is produced by thermal fluctuations or by imperfections of the geometry of the
 channels. While the geometry can be made nearly perfect, thermal fluctuations are inevitable. 
Also, when thinking of experimental measurements of electrons moving on the surface of He in
 micro-channels, we should keep in mind that the experimental methods~\cite{169_ikegami} do not
 allow to control the electron structure itself. 
 The structure is detected indirectly, via measuring the electron current through a narrow
 constriction. 
 Therefore, one cannot judge whether the electrons were perfectly ordered or not before they
 entered the constriction. 
 Of course, the limiting cases of a crystal and liquid can be distinguished by the appearance of
 typical oscillations in the $IV$-curves in case of a Wigner 
 crystal~\cite{169_ikegami,213_rees,214_rees,216_vasylenko,188_vasylenko}. 
 On the other hand, one can expect that weak disorder cannot be detected in the electron current
 measurements, i.e., these measurements cannot distinguish a perfect Wigner crystal from a
 slightly disordered one. 
 However, it is not known whether the order of transitions studied above is sensitive to weak
 disorder. 

\begin{figure} [t!]
\begin{center}
\subfigure{\includegraphics[trim={0.0cm 0 0 0}, scale=0.68]{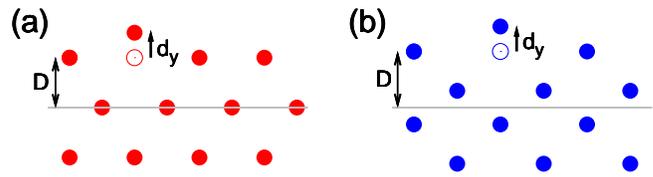}}
\caption{\label{fig:imperfection} (Color online) Schematic representation of the structural
 imperfections induced in the structures of (a) three- and  (b) four-chain configuration. }
\end{center}
\end{figure}

In order to analyze the effect of disorder, we introduce a small displacement of one particle (per
 simulation cell) from its symmetric GS configuration, as shown in Fig.~\ref{fig:imperfection}. 
 Note that in the general case this procedure is equivalent to the effect of non-zero temperature 
 (see, e.g., Ref.~[\onlinecite{003_piacente}]). 
 In particular, a displacement of $10\%$ of the lattice constant from its equilibrium position,
 according to the Lindemann criterion, is treated as (local) melting of the crystal. 

\begin{figure} [t!]
\begin{center}
\subfigure{\includegraphics[trim={0.0cm 0 0 0}, scale=0.68]{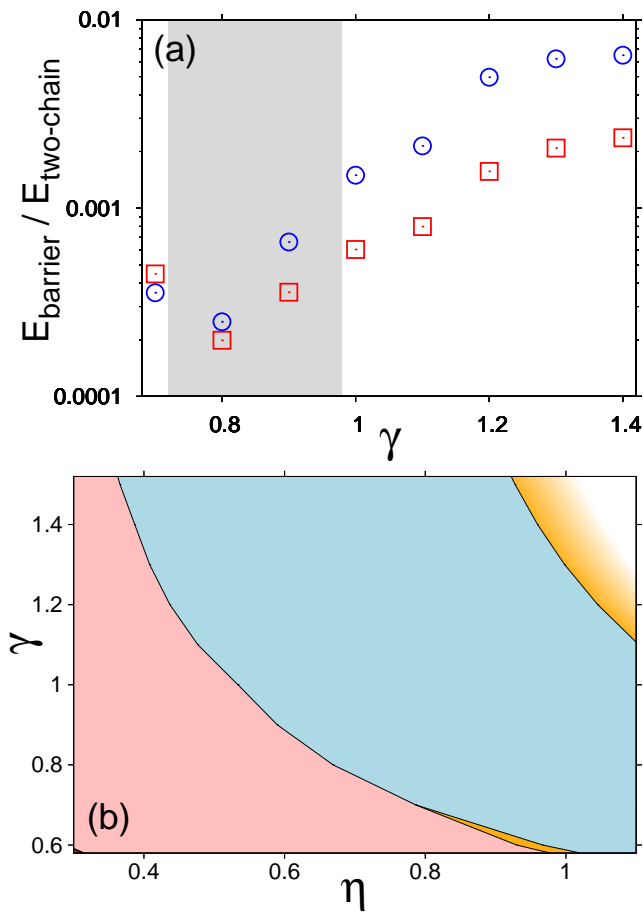}}
\caption{\label{fig:imperf_eb_phd} (Color online) (a) Relative energy barrier as a function of the
 shape parameter ($\gamma$) at the transition point.
 Open circles (squares) represent the energy barrier of transition $2 \rightarrow 3^{*}$
 ($2 \rightarrow 4^{*}$).
 The gray highlighted region indicates the values of $\gamma$ where the transition $2 \rightarrow 4$
 is not stable, as shown in Fig.~\ref{fig:stability}.
 (b) The same phase diagram as in Fig.~\ref{fig:stability}, but with the effect of the induced
 imperfection in the lattice.}
\end{center}
\end{figure}

Following the procedure described above, we increase the particle's density and analyze the
 structural transitions. 
We calculate the relative energy barrier of the transition from two- to the imperfect three-chain
 ($2 \rightarrow 3^{*}$) and to the imperfect four-chain ($2 \rightarrow 4^{*}$) configurations.
 Fig.~\ref{fig:imperf_eb_phd}(a) shows this barrier plotted as a function of $\gamma$ at the
 transition point, circles and squares represent the energy barrier for $2 \rightarrow 3^{*}$ and
 $2 \rightarrow 4^{*}$, respectively. 
One can see that the energy barrier of $2 \rightarrow 3^{*}$ decreases as compared to the one of
 $2 \rightarrow 4^{*}$, resulting in the shrinking of the region where the transition to
 three-chain is more favorable than to the four-chain configuration.
 Therefore, the GS of the system follows the SOT in a wide range of parameters, as clearly shown
 in Figs.~\ref{fig:imperf_eb_phd}(a,b). 
In the highlighted gray region in Fig.~\ref{fig:imperf_eb_phd}(a), the SOT guides the GS due to the instability of the transition
 $2 \rightarrow 4$ configuration, as discussed in previous section.
 However, just a small imperfection, as the one modeled in the present analysis, breaks the
 robustness of the transition $2 \rightarrow 4$, allowing the system to evolve following the SOT
 for a region of $\gamma$ much wider than in case of perfect chain-like structures (cf.
 Fig.~\ref{fig:stability}).
 
Using these results we re-plot the phase diagram for the above studied system ($n=2$ and
 $\kappa=1$) i.e., shown in Fig.~\ref{fig:stability}.
 In the new phase diagram (Fig.~\ref{fig:imperf_eb_phd}(b)) the GS of the system follows the SOT
 for a broad range of densities, starting from $\gamma \approx 0.7$ and higher. 
Therefore, even a small amount of disorder changes dramatically the picture of the ordering of the
 transition in the GS configuration of the system. The non-SOT earlier considered as a ``robust''
 feature (since it survived under various transformations of the confinement potential and the
 inter-particle interaction) turns out to be fragile against a small deformation in the lattice,
 for low particle density.
 This fragility of the non-SOT explains, why experimentally the non-SOT has never been observed
 under real experimental conditions (e.g., non-zero temperature). 

\section{Conclusions} \label{conclusions}

The existence of the so-called non-sequential ordering of structural transitions (non-SOT), when
 increasing the particles density, in a system of repulsive particles confined in a
 quasi-one-dimensional potential was theoretically predicted in a number of works. 
 The non-SOT, when the number of chains follows the sequence 1-2-4-3-4..., was shown to be the
 ``natural'' sequence of transitions for a broad range of the interaction and confinement
 parameters.
 However, in spite of the theoretically predicted robustness of the non-SOT, experiments with
 various interacting particles including electrons on the surface of liquid He in micro-channels,
 colloids in narrow channels, and superconducting vortices in narrow stripes, did not reveal the
 non-SOT. Instead, the transitions followed the SOT, i.e., 1-2-3-4... 

In an attempt to investigate this controversial behavior, we studied in detail the effect of
 boundaries on the sequence of the structural transitions~\cite{211_galvan}. 
 It was shown, in particular, that in case of a Bean-Livingston-type barrier, which appears in
 case of vortices in a superconductor, one observes SOT rather than non-SOT. 
 In this work, we took a deeper insight in this problem, by generalizing our study to
 interparticle interactions of very different range. 
 In addition, we analyzed the effect of fluctuations which provided us with a deeper understanding
 why real systems display SOT rather than the theoretically predicted non-SOT. 

In particular, we investigated the structural transitions for the ground state (GS) of a classical
 system of particles confined in a channel. 
 The inter-particle interaction was modeled as a screened power-law potential, and the profile of
 the channel confining potential was modeled by two different functional forms, both being
 gradually tunable from a parabolic- to a hard-wall-like confinement.
 In this context all the GS configurations are presented by chain-like structures, which
 correspond to Wigner crystal structures. 
 
The effect of the interaction between particles on the ordering of transitions of the GS
 configurations, was investigated for different values of the interaction parameters, varying from
 a short-range to a long-range interaction potential. 
 We analytically calculated the energy of the chain-like structures at zero temperature, and
 analyzed the results obtained for different confinement potentials. 
 We found that the non-SOT is present in all cases studied and it was found irrespective of
 the range of the interaction and for different energy profiles of the confining channel. 

On the other hand, our calculations show that a region where the SOT guides the transitions of the
 GS by increasing particle density, emerges in case of a long-range interaction potential (i.e.,
 when approaching the limit of unscreened Coulomb interaction), and this region is extended for
 longer-range interactions: The longer the range of the interaction, the higher probability to
 find the SOT as a sequence of the GS of the system. 

Two different types of trapping potentials, which control the profile of the channel where the particles
 are confined, were studied: exponential and Gaussian confinement. 
 In both cases the shape parameter allows us to control the shape of the channel, varying it from
 a parabolic-like to a hard-wall-like channel by increasing the shape parameter. 
 We found that for a hard-wall-like confinement, the transitions between different configurations
 occur at lower particle linear density, and the ordering of the transitions is related to the
 range of the interaction potential. 
 Oppositely, in case of a parabolic-like confinement, we found in all studied cases, that the
 transitions between the different phases are always given by a non-SOT. 
 
The existence of the sequential or non-sequential ordering of transitions between phases in the
 GS is determined by the existence or absence of the four-chain GS configuration ``inserted'' in
 the direct transition $2 \rightarrow 3$. 
We investigated the ability of the system to follow either the $2 \rightarrow 3$ transition or
 the $2 \rightarrow 4$, by calculating the energy barrier which the system has to overcome in
 order to reach the final state. 
 We found that, irrespective of the confinement profile and the interaction potential, the barrier
 for the $2 \rightarrow 3$ transition is higher than that for the $2 \rightarrow 4$, thus making
 the non-SOT as the most ``natural'' ordering of transitions for the GS of the system. 

However, when analyzing the stability of the configuration during this transition
 ($2 \rightarrow 3$), we found that if this transition is not allowed as the GS, the reason for
 this behavior is the instability of that state.
 Our numerical results obtained from the analysis of the stability of the states perfectly confirm the
 behavior of the transitions and the phase diagram calculated analytically for this system. 
 The analysis of the stability explains that the robustness of the non-SOT is not just due to the
 energetically favorable arrangement of the particles, but due to the instability of the
 transition $2 \rightarrow 3$. 
 
As further follows from our analysis, the effect of weak fluctuations in the chain-like structure
 can lead to the SOT behavior. 
 Thus our simulations showed that even a small imperfection could change dramatically the ordering
 of transitions facilitating the appearance of the SOT and considerably increasing the window of
 parameters to observe it. 
 This results in a remarkable finding: The robustness of the non-SOT in the system can be
 eliminated by a weak disorder of the Wigner crystal, e.g., due to a non-zero temperature. 

Our results manifested that small fluctuations destroy the intermediate four-chain configuration
 (i.e., the hallmark of the non-SOT), facilitating the experimental observation of the SOT instead
 of the earlier predicted non-SOT as the most ``natural'' sequence of transitions. 
 At the same time, an important result of our analysis is that we found the window of parameters
 where the non-SOT still can be potentially found, even in the presence of weak fluctuations (e.g.,
 for non-zero but relatively low temperature).
 In particular, our analysis showed that the non-SOT is ``protected'' against small
 fluctuations in the regime of high linear density and smooth confinement, therefore, we expect
 that the non-SOT would be experimentally observed in that region, rather than for low density of
 particles confined in a hard-wall channel.
 
\acknowledgments

This work was supported by the Flemish Science Foundation (FWO-Vl) and the Odysseus and
 Methusalem programmes of  the  Flemish  government.
Computational  resources were provided by HPC infrastructure of the University of Antwerp
 (CalcUA) a division of the Flemish Supercomputer Center (VSC).


%

\end{document}